\documentclass[journal,comsoc]{IEEEtran}

\usepackage[T1]{fontenc}
\usepackage{cite}
\usepackage{lipsum}
\usepackage[utf8]{inputenc}
\usepackage[normalem]{ulem}
\usepackage{hyperref}
\usepackage{rotating}
\usepackage{tabularx}
\usepackage{multirow}
\usepackage[table]{xcolor}
\usepackage{url}

\usepackage{adjustbox}
\usepackage{graphicx}
\usepackage{amssymb}
\usepackage{rotating}
\usepackage{tabularx}
\usepackage{url}
\usepackage{rotating}
\usepackage{longtable}
\usepackage{changepage}

\usepackage{ulem}

\usepackage{color}

\ifCLASSINFOpdf
\else
\fi

\usepackage{amsmath}
\usepackage[cmintegrals]{newtxmath}
\usepackage{flushend}

\begin{document}
\title{A Survey of Network-based Intrusion Detection Data Sets}

\author{Markus Ring, Sarah Wunderlich, Deniz Scheuring, Dieter Landes and Andreas Hotho
\thanks{Markus Ring, Sarah Wunderlich, Deniz Scheuring and Dieter Landes were with the Department of Electrical Engineering and Computer Science, Coburg University of Applied Sciences, 96450 Coburg, Germany (e-mail: markus.ring@hs-coburg.de, sarah.wunderlich@hs-coburg.de, deniz.brix@stud.hs-coburg.de, dieter.landes@hs-coburg.de)}
\thanks{Andreas Hotho was with Data Mining and Information Retrieval Group, University of Würzburg, 97074 Würzburg, Germany (e-mail: hotho@informatik.uni-wuerzburg.de)}}

\maketitle

\begin{abstract}
Labeled data sets are necessary to train and evaluate anomaly-based network intrusion detection systems. 
This work provides a focused literature survey of data sets for network-based intrusion detection and describes the underlying packet- and flow-based network data in detail. 
The paper identifies 15 different properties to assess the suitability of individual data sets for specific evaluation scenarios. 
These properties cover a wide range of criteria and are grouped into five categories such as data volume or recording environment for offering a structured search. 
Based on these properties, a comprehensive overview of existing data sets is given. 
This overview also highlights the peculiarities of each data set.
Furthermore, this work briefly touches upon other sources for network-based data such as traffic generators and data repositories.
Finally, we discuss our observations and provide some recommendations for the use and the creation of network-based data sets. 
 
\end{abstract}


\IEEEpeerreviewmaketitle

\section{Introduction}
\label{sec:introduction}

IT security is an important issue and much effort has been spent in the research of intrusion and insider threat detection. 
Many contributions have been published for processing security-related data \cite{chandola2006data,rehak2008camnep,garcia2014empirical,ring2017}, 
detecting botnets~\cite{beigi2014towards,stevanovic2015analysis,wang2017botnet,yingan}, port scans~\cite{staniford2002practical,jung2004fast,sridharan2006connectionless,ring2018detection}, brute force attacks~\cite{sperotto2009hidden,hellemons2012sshcure,javed2013detecting,najafabadi2014machine} and so on. 
All these works have in common that they require representative network-based data sets.
Furthermore, benchmark data sets are a good basis to evaluate and compare the quality of different network intrusion detection systems (NIDS).
Given a labeled data set in which each data point is assigned to the class normal or attack, the number of detected attacks or the number of false alarms may be used as evaluation criteria. 
Unfortunately, there are not too many representative data sets around. 
According to Sommer and Paxson~\cite{sommer2010outside} (2010), the lack of representative publicly available data sets constitutes one of the biggest challenges for anomaly-based intrusion detection. 
Similar statements are made by Malowidzki et al.~\cite{malowidzki2015network} (2015) and Haider et al.~\cite{haider2017generating} (2017). 
However, the community is working on this problem as several intrusion detection data sets have been published over the last years. 
In particular, the Australian Centre for Cyber Security published the UNSW-NB15~\cite{moustafa2015unsw} data set, the University of Coburg published the CIDDS-001~\cite{ring2017flow} data set, or the University of New Brunswick published the CICIDS 2017~\cite{sharafaldin2018toward} data set. 
More data sets can be expected in the future.
However, there is no overall index of existing data sets and it is hard to keep track of the latest developments.

This work provides a literature survey of existing network-based intrusion detection data sets.
At first, the underlying data are investigated in more detail. 
Network-based data appear in packet-based or flow-based format. 
While flow-based data contain only meta information about network connections, packet-based data also contain payload. 
Then, this paper analyzes and groups different data set properties which are often used in literature to evaluate the quality of network-based data sets. 
The main contribution of this survey is an exhaustive literature overview of network-based data sets and an analysis as to which data set fulfills which data set properties. 
The paper focuses on attack scenarios within data sets and highlights relations between the data sets.
Furthermore, we briefly touch upon traffic generators and data repositories as further sources for network traffic besides typical data sets and provide some observations and recommendations. 
As a primary benefit, this survey establishes a collection of data set properties as a basis for comparing available data sets and for identifying suitable data sets, given specific evaluation scenarios. 
Further, we created a website~\footnote{\url{http://www.dmir.uni-wuerzburg.de/datasets/nids-ds}} which references to all mentioned data sets and data repositories and we intend to update this website. 

The rest of the paper is organized as follows. 
The next section discusses related work. 
Section~\ref{sec:data} analyzes packet- and flow-based network data in more detail. 
Section~\ref{sec:properties} discusses typical data set properties which are often used in the literature to evaluate the quality of intrusion detection data sets. 
Section~\ref{sec:datasets} gives an overview of existing data sets and checks each data set against the identified properties of Section~\ref{sec:properties}. 
Section~\ref{sec:further} briefly touches upon further sources for network-based data.  
Observations and recommendations are discussed in Section~\ref{sec:obs} before the paper concludes with a summary.

\section{Related Work}
\label{sec:related}

This section reviews related work on network-based data sets for intrusion detection. 
It should be noted that host-based intrusion detection data sets like ADFA~\cite{creech2013generation} are not considered in this paper. Interested readers may find details on host-based intrusion detection data in Glass-Vanderlan et al.~\cite{glass2018survey}.

Malowidzki et al. \cite{malowidzki2015network} discuss missing data sets as a significant problem for intrusion detection, set up requirements for good data sets, and list available data sets. 
Koch et al.~\cite{koch2014towards} provide another overview of intrusion detection data sets, analyze 13 data sources, and evaluate them with respect to 8 data set properties. 
Nehinbe \cite{nehinbe2011critical} provides a critical evaluation of data sets for IDS and intrusion prevention systems (IPS). 
The author examines seven data sets from different sources (e.g. DARPA data sets and DEFCON data sets), highlights their limitations, and suggests methods for creating more realistic data sets. 
Since many data sets are published in the last four years, we continue previous work \cite{malowidzki2015network, koch2014towards, nehinbe2011critical} from 2011 to 2015, but offer a more up-to-date and more detailed overview than our predecessors.

While many data set papers (e.g., CIDDS-002~\cite{ring2017creation}, ISCX~\cite{shiravi2012toward} or UGR'16~\cite{macia2018ugr}) give just a brief overview of some intrusion detection data sets, Sharafaldin et al. \cite{sharafaldin2017towards} provide a more exhaustive review. 
Their main contribution is a new framework for generating intrusion detection data sets. 
Sharafaldin et al. also analyze 11 available intrusion detection data sets and evaluate them with respect to 11 data set properties. 
In contrast to earlier data set papers, our work focuses on providing a neutral overview of existing network-based data sets rather than contributing an additional data set. 

Other recent papers also touch upon network-based data sets, yet with a different primary focus.
Bhuyan et al. \cite{bhuyan2014network} present a comprehensive review of network anomaly detection.
The authors describe nine existing data sets and analyze data sets which are used by existing anomaly detection methods.  
Similarly, Nisioti et al. \cite{Nisioti2018} focus on unsupervised methods for intrusion detection and briefly refer to 12 existing network-based data sets. 
Yavanoglu and Aydos \cite{yavanoglu2017review} analyze and compare the most commonly used data sets for intrusion detection. 
However, their review contains only seven data sets including other data sets like HTTP CSIC 2010 \cite{gimenez2010http}.  
All in all, these works tend to have different research objectives and only touch upon network-based data sets marginally.

\section{Data}
\label{sec:data}

Normally, network traffic is captured either in packet-based or flow-based format. 
Capturing network traffic on packet-level is usually done by mirroring ports on network devices. 
Packet-based data encompass complete payload information.
Flow-based data are more aggregated and usually contain only metadata from network connections. 
Wheelus et al. highlight the distinction through an illustrative comparison: 
\textit{"A good example of the difference between captured packet inspection and NetFlow would be viewing a forest by hiking through the forest as opposed to flying over the forest in a hot air balloon"}~\cite{wheelus2014session}.
In this work, a third category (\textit{other} data) is introduced. 
The category \textit{other} has no standard format and varies for each data set.

\subsection{Packet-based data}

\begin{figure}[]
	\centering
	\includegraphics[width=0.4\textwidth]{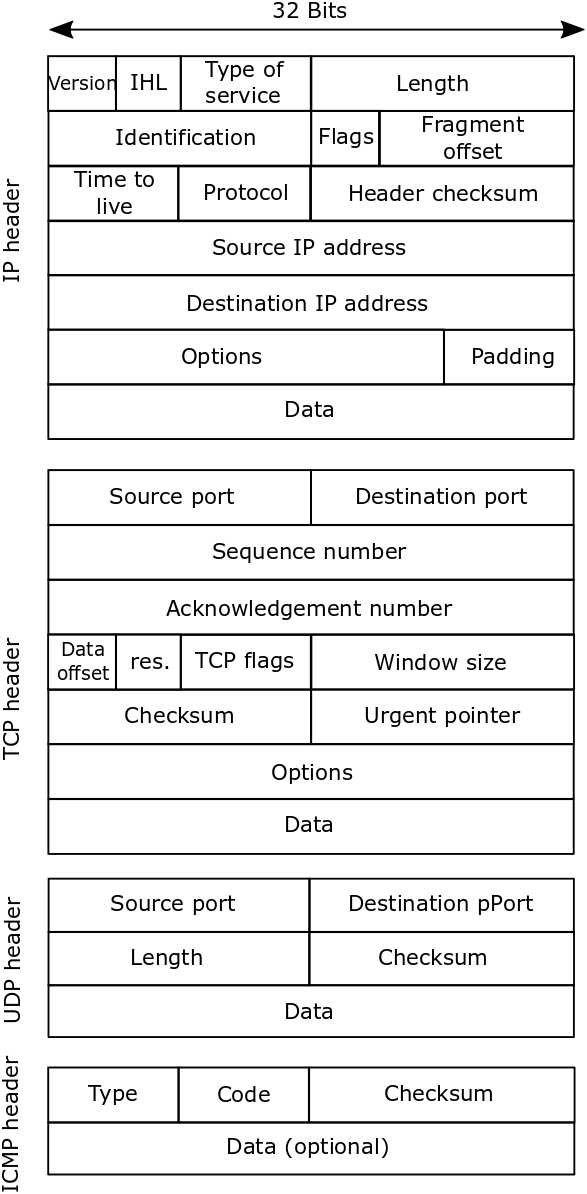}
	\caption{
		IP, TCP, UDP and ICMP header after \cite{tanenbaum2002computer}.
	}
	\label{fig:tcpip}
\end{figure}

Packet-based data is commonly captured in pcap format and contains payload.
Available metadata depends on the used network and transport protocols. 
There are many different protocols and the most important ones being TCP, UDP, ICMP and IP. 
Figure \ref{fig:tcpip} illustrates the different headers. 
TCP is a reliable transport protocol and encompasses metadata like sequence number, acknowledgment number, TCP flags, or checksum values.
UDP is a connection-less transport protocol and has a smaller header than TCP which contains only four fields, namely source port, destination port, length and checksum.
In contrast to TCP and UDP, ICMP is a supporting protocol containing status messages and is thus even smaller.
Normally, there is also an IP header available beside the header of the transport protocol. 
The IP header provides information such as source and destination IP addresses and is also shown in Figure~\ref{fig:tcpip}. 

\subsection{Flow-based data}
Flow-based network data is a more condensed format which contains mainly meta information about network connections.
Flow-based data aggregate all packets which share some properties within a time window into one flow and usually do not include any payload. 
The default five-tuple definition, i.e., source IP address, source port, destination IP address, destination port and transport protocol \cite{claise2008specification}, is a widely used standard for matching properties in flow-based data. 
Flows can appear in unidirectional or bidirectional format. 
The unidirectional format aggregates all packets from host $A$ to host $B$ which share the above mentioned properties into one flow.
All packets from host $B$ to host $A$ are aggregated into another unidirectional flow. 
In contrast, a bidirectional flow summarizes all packets between hosts $A$ and $B$, regardless of direction. 

\begin{table}[]
	
	\caption{
		Attributes in Flow-based Network Traffic. 
	}
	\label{tbl:netflow}
	\centering
	\begin{tabular}{p{1em}p{14em}}
		\hline\noalign{\smallskip}
		\# & Attribute \\
		\hline
		1 & Date first seen  \\
		2 & Duration \\
		3 & Transport protocol \\
		4 & Source IP address \\
		5 & Source port \\
		6 & Destination IP address \\
		7 & Destination port\\
		8 & Number of transmitted bytes\\
		9 & Number of transmitted packets \\
		10 & TCP flags \\
	\end{tabular}
	
\end{table}

Typical flow-based formats are NetFlow \cite{claise2004cisco}, IPFIX \cite{claise2008specification}, sFlow \cite{phaal2004} and OpenFlow \cite{mckeown2008openflow}. 
Table \ref{tbl:netflow} gives an overview of typical attributes within flow-based network traffic. 
Depending on the specific flow format and flow exporter, additional attributes like bytes per second, bytes per packet, TCP flags of the first packet, or even the calculated entropy of the payload can be extracted. 

Furthermore, it is possible to convert packet-based data to flow-based data (but not vice versa) with tools like nfdump\footnote{\url{https://github.com/phaag/nfdump}} or YAF\footnote{\url{https://tools.netsa.cert.org/yaf/}}.  
Readers interested in the differences between flow exporters may find additional details in~\cite{haddadi2016benchmarking}, together with an analysis of how different flow exporters affect botnet classification.

\subsection{Other data}
\label{sec:other}
This category includes all data sets that are neither purely packet-based nor flow-based.  
An example of this category might be flow-based data sets which have been enriched with additional information from packet-based data or host-based log files.
The KDD CUP 1999 \cite{kddcup} data set is a well-known representative of this category. 
Each data point has network-based attributes like the number of transmitted source bytes or TCP flags, but has also host-based attributes like number of failed logins.
As a consequence, each data set of this category has its own set of attributes.  
Since each data set must be analyzed individually, we do not make any general statements about available attributes.

\section{Data Set Properties}
\label{sec:properties}

To be able to compare different intrusion detection data sets side by side and to help researchers finding appropriate data sets for their specific evaluation scenario, it is necessary to define common properties as evaluation basis.
Therefore, we explore typical data set properties that are used in the literature to assess intrusion detection data sets.
The general concept FAIR~\cite{wilkinson2016fair} defines four principles that scholarly data should fulfill, namely \textit{Findability}, \textit{Accessibility}, \textit{Interoperability} and \textit{Reusability}.
While concurring with this general concept, this work uses more detailed data set properties to provide a focused comparison of network-based intrusion detection data sets.
Generally, different data sets emphasize different data set properties.
For instance, the UGR'16 data set \cite{macia2018ugr} emphasizes a long recording time to capture periodic effects while the ISCX data set \cite{shiravi2012toward} focuses on accurate labeling.
Since we aim at investigating more general properties for network-based intrusion detection data sets, we try to unify and generalize properties used in literature rather than adopting all of them.
For example, some approaches evaluate the presence of specific kind of attacks like DoS (Denial of Service) or Browser injections.
The presence of certain attack types may be a relevant property for evaluating detection approaches for those specific attack types, but are meaningless for other approaches.
Hence, we use the general property attacks which describes the presence of malicious network traffic (see Table \ref{tbl:datasets}).
Section~\ref{sec:datasets} provides more details on the different attack types in the data sets together with a discussion of other particular properties.

\begin{figure*}[]
	\centering
	\includegraphics[width=0.7\textwidth]{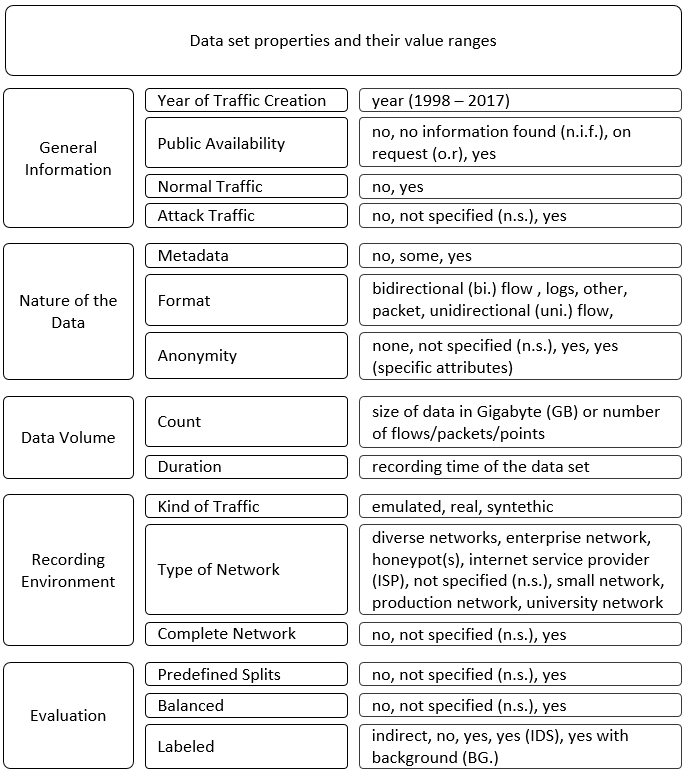}
	\caption{
		Data set properties and their value ranges.
	}
	\label{fig:properties}
\end{figure*}

We do not develop an evaluation score like Haider et al.~\cite{haider2017generating} or Sharafaldin et al.~\cite{sharafaldin2017towards} since we do not want to judge the importance of different data set properties. 
In our opinion, the importance of certain properties depends on the specific evaluation scenario and should not be generally judged in a survey. 
Rather, readers should be put in a position to find suitable data sets for their needs. 
Therefore, we group the data set properties discussed below in five categories to support systematic search. 
Figure~\ref{fig:properties} summarizes all data set properties and their value ranges.

\subsection{General Information}
The following four properties reflect general information about the data set, namely the year of creation, availability, presence of normal and malicious network traffic.

\subsubsection{Year of Creation}
Since network traffic is subject to concept drift and new attack scenarios appear daily, the age of an intrusion detection data set plays an important role.
This property describes the year of creation. 
The year in which the underlying network traffic of a data set was captured is more relevant for up-to-dateness than the year of its publication.

\subsubsection{Public Availability}
Intrusion detection data sets should be publicly available to serve as a basis for comparing different intrusion detection methods. 
Furthermore, the quality of data sets can only be checked by third parties if they are publicly available. 
Table \ref{tbl:datasets} encompasses three different characteristics for this property: yes, o.r. (on request), and no. 
On request means that access will be granted after sending a message to the authors or the responsible person. 

\subsubsection{Normal User Behavior}
This property indicates the availability of normal user behavior within a data set and takes the values yes or no. 
The value yes indicates that there is normal user behavior within the data set, but it does not make any statements about the presence of attacks.
In general, the quality of an IDS is primarily determined by its attack detection rate and false alarm rate. 
Therefore, the presence of normal user behavior is indispensable for evaluating an IDS. 
The absence of normal user behavior, however, does not make a data set unusable, but rather indicates that it has to be merged with other data sets or with real world network traffic.
Such a merging step is often called overlaying or salting \cite{aviv2011challenges, celik2011salting}.

\subsubsection{Attack Traffic}
IDS data sets should include various attack scenarios. 
This property indicates the presence of malicious network traffic within a data set and has the value yes if the data set contains at least one attack. 
Table~\ref{tbl:attacks} provides additional information about the specific attack types.

\subsection{Nature of Data}
Properties of this category describe the format of the data sets and the presence of meta information. 

\subsubsection{Metadata}
Content-related interpretation of packet-based and flow-based network traffic is difficult for third parties. 
Therefore, data sets should come along with metadata to provide additional information about the network structure, IP addresses, attack scenarios and so on. 
This property indicates the presence of additional metadata.

\subsubsection{Format}
Network intrusion detection data sets appear in different formats. 
We roughly divide them into three formats (see Section \ref{sec:data}).
(1) Packet-based network traffic (e.g. pcap) contains network traffic with payload.  
(2) Flow-based network traffic (e.g. NetFlow) contains only meta information about network connections. 
(3) Other types of data sets may contain, e.g., flow-based traces with additional attributes from packet-based data or even from host-based log files. 

\subsubsection{Anonymity}
Frequently, intrusion detection data sets may not be published due to privacy reasons or are only available in anonymized form. 
This property indicates if data is anonymized and which attributes are affected. 
The value none in Table~\ref{tbl:datasets} indicates that no anonymization has been performed. 
The value yes (IPs) means that IP addresses are either anonymized or removed from the data set. 
Similarly, yes (payload) indicates that payload information is anonymized or removed from packet-based network traffic.

\subsection{Data Volume}
Properties in this category characterize data sets in terms of volume and duration.  

\subsubsection{Count}
The property count describes a data set's size as either the number of contained packets/flows/points or the physical size in Gigabyte (GB). 

\subsubsection{Duration}
Data sets should cover network traffic over a long time for capturing periodical effects (e.g., daytime vs. night or weekday vs. weekend) \cite{macia2018ugr}.
The property duration provides the recording time of each data set.

\subsection{Recording Environment}
Properties in this category delineate the network environment and conditions in which the data sets are captured. 

\subsubsection{Kind of Traffic}
The property Kind of Traffic describes three possible origins of network traffic: real, emulated, or synthetic. 
Real means that real network traffic was captured within a productive network environment. 
Emulated means that real network traffic was captured within a test bed or emulated network environment. 
Synthetic means that the network traffic was created synthetically (e.g., through a traffic generator) and not captured by a real (or virtual) network device.

\subsubsection{Type of Network}
Network environments in small and medium-sized companies are fundamentally different from internet service providers (ISP). 
As a consequence, different environments require different security systems and evaluation data sets should be adapted to the specific environment. 
This property describes the underlying network environment in which the respective data set was created. 

\subsubsection{Complete Network}
This property is adopted from Sharafaldin et al. \cite{sharafaldin2017towards} and indicates if the data set contains the complete network traffic from a network environment with several hosts, router and so on.  
If the data set contains only network traffic from a single host (e.g., honeypot) or only some protocols from the network traffic (e.g., exclusively SSH traffic), the value is set to no.

\subsection{Evaluation}
The following properties are related to the evaluation of intrusion detection methods using network-based data sets.
More precisely, the properties denote the availability of predefined subsets, the data set's balance, and the presence of labels.

\subsubsection{Predefined Splits}
Sometimes it is difficult to compare the quality of different IDS, even if they are evaluated on the same data set. 
In that case, it must be clarified whether the same subsets are used for training and evaluation. 
This property provides the information if a data set comes along with predefined subsets for training and evaluation. 

\subsubsection{Balanced}
Often, machine learning and data mining methods are used for anomaly-based intrusion detection. 
In the training phase of such methods (e.g., decision tree classifiers), data sets should be balanced with respect to their class labels. 
Consequently, data sets should contain the same number of data points from each class (normal and attack).
Real-world network traffic, however, is not balanced and contains more normal user behavior than attack traffic. 
This property indicates if data sets are balanced with respect to their class labels. 
Imbalanced data sets should be balanced by appropriate preprocessing before data mining algorithms are used. 
He and Garcia \cite{he2009learning} provide a good overview of learning from imbalanced data.

\subsubsection{Labeled}
Labeled data sets are necessary for training supervised methods and for evaluating supervised as well as unsupervised intrusion detection methods. 
This property denotes if data sets are labeled or not. 
If there are at least the two classes normal and attack, this property is set to yes. 
Possible values in this property are: yes, yes with BG. (yes with background), yes (IDS), indirect, and no. 
Yes with background means that there is a third class background. 
Packets, flows, or data points which belong to the class background could be normal or attack. 
Yes (IDS) means that some kind of intrusion detection system was used to create the data set's labels.  
Some labels of the data set might be wrong since an IDS might be imperfect. 
Indirect means that the data set has no explicit labels, but labels can be created on one's own from additional log files.

\section{Data Sets}
\label{sec:datasets}

In our opinion, the data set properties \textit{Labeled} and \textit{Format} are the most decisive properties when searching for adequate network-based data sets.
The intrusion detection method (supervised or unsupervised) determines if labels are necessary or not and which kind of data is required (packet, flow or \textit{other}).
Therefore, Table~\ref{tbl:decision} provides a classification of all investigated network-based data sets with respect to these two properties. 
A more detailed overview of network-based intrusion detection data sets with respect to the data set properties of Section~\ref{sec:properties} is given in Table~\ref{tbl:datasets}. 
The presence of specific attack scenarios is an important aspect when searching for a network-based data set. 
Therefore, Table~\ref{tbl:datasets} indicates the presence of attack traffic while Table~\ref{tbl:attacks} provides details on specific attacks within a data set.
Papers on data sets describe attacks on different abstraction levels.  
Vasudevan et al. \cite{vasudevan}, for instance,  characterized attack traffic within their data set (SSENET-2011) as follows: \textit{"Nmap, Nessus, Angry IP scanner, Port Scanner, Metaploit, Backtrack OS, LOIC, etc., were some of the attack tools used by the participants to launch the attacks."}. In contrast, Ring et al. specify the number and different types of executed port scans in their CIDDS-002 data set \cite{ring2017creation}.
Consequently, the abstraction level of attack descriptions may vary in Table~\ref{tbl:attacks}.
A detailed description of all attack types is beyond the scope of this work. 
Rather, we refer interested readers to the open access paper \textit{"From Intrusion Detection to an Intrusion Response System: Fundamentals, Requirements, and Future Directions"} by Anwar et al. \cite{anwar2017intrusion}.

\begin{figure}[]
	\centering
	\includegraphics[width=0.47\textwidth]{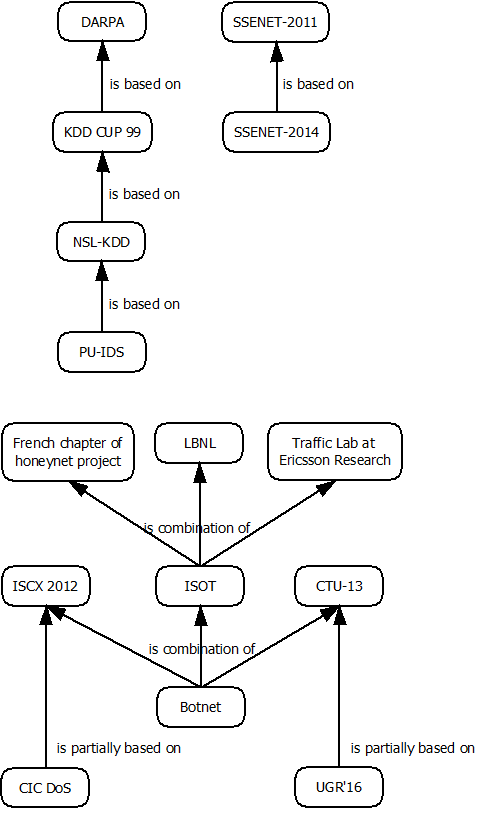}
	\caption{
		Relationships between the data sets in Table \ref{tbl:datasets}.
	}
	\label{fig:relations}
\end{figure}

Further, some data sets are modifications or combinations of others.
Figure \ref{fig:relations} shows the interrelationships among several well-known data sets. 

\begin{table}[]

	\caption{
		Decision support table for finding appropriate network-based data sets. 
		Some data sets like CTU-13 provide several data formats and appear in several columns. 
		(+) indicates that the data set is publicly available. 
		(?) indicates that we were not able to find the data set.
		(-) indicates that the data set is not publicly available.
	}
	%\rowcolors{2}{gray!25}{white}
	\label{tbl:decision}
	\centering
	\resizebox{0.485\textwidth}{!}{
	\begin{tabular}{p{5em}||p{7.5em}|p{7.5em}|p{7.5em}}
		& \multicolumn{3}{c}{Format}  \\
		\hline	
		\hline
		Labeled & packet & flow & other \\
		\hline
		yes & (+) Botnet 		& (+) CICIDS 2017 	& (+) AWID \\
		& (+) CIC DoS  		& (+) CIDDS-001 	& (+) KDD CUP 99 \\
		& (+) CICIDS 2017 		& (+) CIDDS-002 	& (+) Kyoto 2006+\\
		& (+) DARPA 			& (+) ISCX 2012 	& (+) NSL-KDD \\
		& (+) DDoS 2016 		& (+) TUIDS 	& (+) UNSW-NB 15 \\
		& (+) ISCX 2012 		& (+) Twente 		& 	\\
		& (+) ISOT 			&  		&  \\
		& (+) NDSec-1 		&   				& \\
		& (+) NGIDS-DS 	& & \\
		& (+) TRAbID 		& & \\
		& (+) TUIDS 		& & \\
		& (+) UNSW-NB15 	& & \\
		&					& & \\
		& 					&					& (?) PU-IDS \\
		& 					&					& (?) SSENET-2011  \\
		& 					&					& (?) SSENET-2014  \\
		&					& & \\
		& (-) IRSC 		& (-) IRSC 			& (-) SANTA \\
		
		\hline
		yes with BG & (+) CTU-13 & (+) CTU-13 & \\
		&		 & (+) UGR'16 & \\
		\hline
		yes (IDS) & & (?) PUF &  \\
		\hline
		indirect & & (+) SSHCure &  \\
		\hline
		no & (+) Booters  	& (+) Kent 2016 & \\ 
		& (+) CDX 		& (+) UNIBS & \\ 
		& (+) LBNL 		& (+) Unified Host and Network & \\

	\end{tabular}}

\end{table}

\begin{sidewaystable*}[htbp!]
	
	\caption{
		Overview of Network-based Data Sets.  
	}
	%\rowcolors{2}{gray!25}{white}
	\label{tbl:datasets}
	\centering
	%\begin{tabular}{p{1em}p{8em}p{2em}p{2em}p{3em}p{3em}p{3em}p{2em}p{2em}p{2em}p{2em}p{2em}p{2em}p{2em}p{2em}p{2em}p{2em}p{2em}}
	
	\rowcolors{2}{gray!25}{white}
	\begin{adjustwidth}{-0.5cm}{}	
		\resizebox{1.0\textwidth}{!}{
			\begin{tabular}{p{8.2em}!{\vrule width 2pt}p{5.5em}|p{3em}|p{3em}|p{3em}!{\vrule width 2pt}p{2.5em}|p{7em}|p{5.8em}!{\vrule width 2pt}p{6.5em}|p{5em}!{\vrule width 2pt}p{4em}|l|p{3.7em}!{\vrule width 2pt}p{2.5em}|p{3.3em}|p{4em}}
				
				& \multicolumn{4}{c!{\vrule width 2pt}}{\textbf{General Information}} & \multicolumn{3}{c!{\vrule width 2pt}}{\textbf{Nature of the Data}} & \multicolumn{2}{c!{\vrule width 2pt}}{\textbf{Data Volume}} & \multicolumn{3}{c!{\vrule width 2pt}}{\textbf{Recording Environment}} & \multicolumn{3}{c}{\textbf{Evaluation}}  \\ 
				%{p{8em}!{\vrule width 2pt}p{6em}|p{3em}|p{5em}|l|p{8em}!{\vrule width 2pt}l|p{2.5em}|p{6em}!{\vrule width 2pt}p{5em}|p{6.5em}|p{4em}!{\vrule width 2pt}p{2.5em}|p{2.5em}!{\vrule width 2pt}p{2.5em}|p{3em}}
				\hline
				Data Set
				&   Year of Traffic Creation
				&   Public Avail.
				&   Normal Traffic 
				&   Attack Traffic 
				&   Meta-data
				&   Format
				&   Anonymity
				&   Count
				&   Duration
				&   Kind of Traffic
				&   Type of Network
				&   Compl. Network
				&   Predef. Splits
				&   Balanced 
				&   Labeled \\
				\hline
				AWID \cite{kolias} & 2015 & o.r. & yes & yes  & yes & other & none & 37M packets & 1 hour & emulated & small network & yes & yes & no & yes \\
				Booters \cite{booters} &2013& yes & no & yes & no & packet & yes & 250GB packets & 2 days & real & small network & no & no & no & no\\
				Botnet \cite{beigi2014towards} &2010/2014& yes & yes & yes & yes & packet & none & 14GB packets & n.s. & emulated & diverse networks & yes & yes & no & yes \\
				CIC DoS \cite{jazi2017detecting} &2012/2017& yes & yes & yes  & no & packet & none & 4.6GB packets & 24 hours & emulated & small network & yes & no & no & yes \\
				CICIDS 2017 \cite{sharafaldin2018toward} &2017& yes & yes & yes & yes & packet, bi. flow & none & 3.1M flows & 5 days & emulated & small network & yes & no & no & yes \\
				CIDDS-001 \cite{ring2017flow} &2017& yes & yes & yes  & yes & uni. flow & yes (IPs) & 32M flows & 28 days & emulated and real & small network & yes & no & no & yes \\
				CIDDS-002 \cite{ring2017creation} &2017& yes & yes & yes  & yes & uni. flow & yes (IPs) & 15M flows & 14 days & emulated & small network & yes & no & no & yes\\
				CDX \cite{sangster2009toward} &2009& yes & yes & yes & yes & packet & none & 14GB packets & 4 days & real & small network & yes & no & no & no  \\
				CTU-13 \cite{garcia2014empirical} &2013& yes & yes & yes  & yes & uni. and bi. flow, paket & yes (payload) & 81M flows & 125 hours & real & university network & yes & no & no & yes with BG. \\
				DARPA \cite{lippmann2000evaluating,lippmann20001999}  & 1998/99 & yes & yes & yes  & yes & packet, logs & none & n.s. & 7/5 weeks & emulated & small network & yes & yes & no & yes \\
				DDoS 2016 \cite{alkasassbeh2016detecting} &2016& yes & yes & yes & no & packet & yes (IPs) & 2.1M packets & n.s. & synthetic & n.s. & n.s. & no & no & yes \\
				IRSC \cite{zuech2015new} &2015& no & yes & yes  & no & packet, flow & n.s. & n.s. & n.s. & real & production network & yes & n.s. & n.s. & yes\\
				ISCX 2012 \cite{shiravi2012toward} &2012& yes & yes & yes  & yes & packet, bi. flow & none & 2M flows & 7 days & emulated & small network & yes & no & no & yes\\
				ISOT \cite{saad2011detecting} & 2010& yes & yes & yes  & yes & packet & none & 11GB packets & n.s. & emulated & small network & yes & no & no & yes \\
				KDD CUP 99 \cite{kddcup} &1998& yes & yes & yes  & no & other & none & 5M points & n.s. & emulated & small network & yes & yes & no & yes\\
				Kent 2016 \cite{kent-2015-cyberdata1,akent-2015-enterprise-data} &2016& yes & yes & n.s.  & no & uni. flow, logs & yes (IPs, Ports, date) & 130M flows & 58 days & real & enterprise network & yes & no & no & no\\ 
				Kyoto 2006+ \cite{song2011statistical} & 2006 to 2009 & yes & yes & yes  & no & other & yes (IPs) & 93M points & 3 years & real & honeypots & no & no & no & yes\\
				LBNL \cite{pang2005} & 2004 / 2005 & yes & yes & yes  & no & packet & yes & 160M packets & 5 hours & real & enterprise network & yes & no & no & no \\
				NDSec-1 \cite{beer2017new} &2016& o.r. & no & yes  & no & packet, logs & none & 3.5M packets & n.s. & emulated & small network & yes & no & no & yes\\
				NGIDS-DS \cite{haider2017generating} &2016& yes & yes & yes  & no & packet, logs & none & 1M packets & 5 days & emulated & small network & yes & no & no & yes\\
				NSL-KDD \cite{tavallaee2009analysis} &1998& yes &  yes & yes  & no & other & none &  150k points & n.s. & emulated & small network & yes & yes & no & yes \\
				PU-IDS \cite{singh2015reference} &1998& n.i.f. & yes & yes  & no & other & none & 200k points & n.s. & synthetic & small network & yes & no & no & yes \\
				PUF \cite{sharma2018} &2018& n.i.f. & yes & yes  & no & uni. flow & yes (IPs) & 300k flows & 3 days & real & university network & no & no  & no & yes (IDS) \\
				SANTA \cite{wheelus2014session} &2014& no & yes  & yes & no & other & yes (payload) & n.s. & n.s. & real & ISP & yes & n.s. & no & yes \\
				SSENET-2011 \cite{vasudevan} &2011& n.i.f. & yes & yes  & no & other & none & n.s. & 4 hours & emulated & small network & yes & no & no & yes\\
				SSENET-2014 \cite{bhattacharya2014ssenet} &2011& n.i.f. & yes & yes  & no & other & none & 200k points & 4 hours & emulated & small network & yes & yes & yes & yes\\
				SSHCure \cite{hofstede2014ssh} & 2013 / 2014 & yes & yes & yes  & no & uni. and bi. flow, logs & yes (IPs) & 2.4GB flows (compressed) & 2 months & real & university network & yes & no & no & indirect\\
				TRAbID \cite{viegas2017toward} &2017& yes & yes & yes  & no & packet & yes (IPs) & 460M packets & 8 hours & emulated & small network & yes & yes & no & yes\\
				TUIDS \cite{gogoi2012packet}, \cite{bhuyan2015towards} & 2011 / 2012 & o.r. & yes & yes  & no & packet, bi. flow & none & 250k flows & 21 days & emulated & medium network & yes & yes & no & yes \\
				Twente \cite{sperotto2009labeled} &2008& yes & no & yes & yes & uni. flow & yes (IPs) & 14M flows & 6 days & real & honeypot & no & no & no & yes \\
				UGR'16 \cite{macia2018ugr} &2016& yes & yes & yes  & some & uni. flows & yes (IPs) & 16900M flows & 4 months & real & ISP & yes & yes & no & yes with BG. \\
				UNIBS \cite{gringoli2009gt} &2009& o.r. & yes & no  & no & flow & yes (IPs) & 79k flows & 3 days & real & university network & yes & no & no & no \\
				Unified Host and Network \cite{turcotte2017unified} &2017& yes & yes & n.s.  & no & bi. flows, logs & yes (IPs and date) & 150GB flows (compressed) & 90 days & real & enterprise network & yes & no & no & no\\
				UNSW-NB15 \cite{moustafa2015unsw} &2015& yes & yes & yes  & yes & packet, other & none & 2M points & 31 hours & emulated & small network & yes & yes & no & yes\\
				
		\end{tabular}}
		\small{\newline\newline\newline}
		\small{n.s. = not specified, n.i.f. = no information found, uni. flow = unidirectional flow, bi. flow = bidirectional flow, yes with BG. = yes with background labels}
	\end{adjustwidth}
\end{sidewaystable*}

\begin{table}[]
	\caption{
		Attacks within the network-based data sets of Table \ref{tbl:datasets}.   
		Specific attack information (e.g. name of the executed botnet) and used tools are provided in round brackets if available. 
	}
	%\rowcolors{2}{gray!25}{white}
	\label{tbl:attacks}
	\rowcolors{2}{gray!25}{white}
	\centering
	\begin{adjustwidth}{0cm}{0cm}
		%\resizebox{0.8\textwidth}{!}{%
			\begin{tabular}{p{8.2em}|p{20em}}
				Data Set & Attacks \\
				\hline
				AWID \cite{kolias}& Popular attacks on 802.11 (e.g. authentication request, ARP flooding, injection, probe request) \\
				Booters \cite{booters} & 9 different DDoS attacks \\
				Botnet \cite{beigi2014towards}& botnets (Menti, Murlo, Neris, NSIS, Rbot, Sogou, Strom, Virut, Zeus) \\
				CIC DoS \cite{jazi2017detecting} & Application layer DoS attacks (executed through ddossim, Goldeneye, hulk, RUDY, Slowhttptest, Slowloris)\\
				CICIDS 2017 \cite{sharafaldin2018toward} & botnet (Ares), cross-site-scripting, DoS (executed through Hulk, GoldenEye, Slowloris, and Slowhttptest), DDoS (executed through LOIC), heartbleed, infiltration, SSH brute force, SQL injection\\
				CIDDS-001 \cite{ring2017flow} & DoS, port scans (ping-scan, SYN-Scan), SSH brute force \\
				CIDDS-002 \cite{ring2017creation} & port scans (ACK-Scan, FIN-Scan, ping-Scan, UDP-Scan, SYN-Scan) \\
				CDX \cite{sangster2009toward} & not specified \\
				CTU-13 \cite{garcia2014empirical} & botnets (Menti, Murlo, Neris, NSIS, Rbot, Sogou, Virut)  \\
				DARPA \cite{lippmann2000evaluating},\cite{lippmann20001999} & DoS, privilege escalation (remote-to-local and user-to-root), probing  \\
				DDoS 2016 \cite{alkasassbeh2016detecting} & DDoS (HTTP flood, SIDDOS, smurf ICMP flood, UDP flood) \\
				IRSC \cite{zuech2015new} & n.s.  \\	
				ISCX 2012 \cite{shiravi2012toward} & Four attack scenarios (1: Infiltrating the network from the inside; 2: HTTP DoS; 3: DDoS using an IRC botnet; 4: SSH brute force) \\	
				ISOT \cite{saad2011detecting} & botnet (Storm, Waledac) \\
				KDD CUP 99 \cite{kddcup} & DoS, privilege escalation (remote-to-local and user-to-root), probing \\
				Kent 2016 \cite{kent-2015-cyberdata1}, \cite{akent-2015-enterprise-data} & not specified \\ 
				Kyoto 2006+ \cite{song2011statistical} & Various attacks against honeypots (e.g. backscatter, DoS, exploits, malware, port scans, shellcode) \\
				LBNL \cite{pang2005} & port scans \\
				NDSec-1 \cite{beer2017new} & botnet (Citadel), brute force (against FTP, HTTP and SSH), DDoS (HTTP floods, SYN flooding and UDP floods), exploits, probe, spoofing, SSL proxy, XSS/SQL injection \\
				NGIDS-DS \cite{haider2017generating} & backdoors, DoS, exploits, generic, reconnaissance, shellcode, worms \\
				NSL-KDD \cite{tavallaee2009analysis} & DoS, privilege escalation (remote-to-local and user-to-root), probing   \\
				PU-IDS \cite{singh2015reference} & DoS, privilege escalation (remote-to-local and user-to-root), probing  \\
				PUF \cite{sharma2018}  & DNS attacks \\
				SANTA \cite{wheelus2014session} & (D)DoS (ICMP flood, RUDY, SYN flood), DNS amplification, heartbleed, port scans  \\
				SSENET-2011 \cite{vasudevan} & DoS (executed through LOIC), port scans (executed through Angry IP Scanner, Nessus, Nmap), various attack tools (e.g. metasploit) \\
				SSENET-2014 \cite{bhattacharya2014ssenet} & botnet, flooding, privilege escalation, port scans\\
				SSHCure \cite{hofstede2014ssh} & SSH attacks \\
				TRAbID \cite{viegas2017toward} & DoS (HTTP flood, ICMP flood, SMTP flood, SYN flood, TCP keepalive), port scans (ACK-Scan, FIN-Scan, NULL-Scan, OS Fingerprinting, Service Fingerprinting, UDP-Scan, XMAS-Scan)  \\
				TUIDS \cite{gogoi2012packet}, \cite{bhuyan2015towards} & botnet (IRC), DDoS (Fraggle flood, Ping flood, RST flood, smurf ICMP flood, SYN flood, UDP flood), port scans (e.g. FIN-Scan, NULL-Scan, UDP-Scan, XMAS-Scan), coordinated port scan, SSH brute force\\
				Twente \cite{sperotto2009labeled} & Attacks against a honeypot with three open services (FTP, HTTP, SSH) \\
				UGR'16 \cite{macia2018ugr} & botnet (Neris), DoS, port scans, SSH brute force, spam \\
				UNIBS \cite{gringoli2009gt} & none \\
				Unified Host and Network \cite{turcotte2017unified} & not specified \\
				UNSW-NB15 \cite{moustafa2015unsw} & backdoors, DoS, exploits, fuzzers, generic, port scans, reconnaissance, shellcode, spam, worms\\		
		\end{tabular}%}
	\end{adjustwidth}
	
\end{table}

\subsection*{Network-based data sets in alphabetical order}

\textbf{AWID \cite{kolias}}. 
AWID is a publicly available data set\footnote{\url{http://icsdweb.aegean.gr/awid/index.html}} which is focused on 802.11 networks.  
Its creators used a small network environment (11 clients) and captured WLAN traffic in packet-based format. 
In one hour, 37 million packets were captured. 156 attributes are extracted from each packet. 
Malicious network traffic was generated by executing 16 specific attacks against the 802.11 network. 
AWID is labeled and split into a training and a test subset.

\textbf{Booters \cite{booters}}. 
Booters are Distributed Denial of Service (DDoS) attacks offered as a service by criminals. 
Santanna et. al \cite{booters} published a data set which includes traces of nine different booter attacks which were executed against a null-routed IP address within their network environment. 
The resulting data set is recorded in packet-based format and consists of more than 250GB of network traffic.  
Individual packets are not labeled, but the different booter attacks are split into different files. 
The data set is publicly available\footnote{\url{https://www.simpleweb.org/wiki/index.php}}, but names of booters are anonymized for privacy reasons.

\textbf{Botnet \cite{beigi2014towards}}. 
The Botnet data set is a combination of existing data sets and is publicly available\footnote{\url{http://www.unb.ca/cic/datasets/botnet.html}}. 
The creators of Botnet used the overlay methodology of \cite{aviv2011challenges} to combine (parts of) the ISOT \cite{saad2011detecting}, ISCX 2012 \cite{shiravi2012toward} and CTU-13 \cite{garcia2014empirical} data sets. 
The resulting data set contains various botnets and normal user behavior. 
The Botnet data set is divided into a 5.3 GB training subset  and a 8.5 GB test subset, both in packet-based format.

\textbf{CIC DoS \cite{jazi2017detecting}}. 
CIC DoS is a data set from the Canadian Institute for Cybersecurity and is publicly available\footnote{\url{http://www.unb.ca/cic/datasets/dos-dataset.html}}. 
The authors' intention was to create an intrusion detection data set with application layer DoS attacks. 
Therefore, the authors executed eight different DoS attacks on the application layer.
Normal user behavior was generated by combining the resulting traces with attack-free traffic from the ISCX 2012 \cite{shiravi2012toward} data set. 
The resulting data set is available in packet-based format and contains 24 hours of network traffic.

\textbf{CICIDS 2017 \cite{sharafaldin2018toward}}. 
CICIDS 2017 was created within an emulated environment over a period of 5 days and contains network traffic in packet-based and bidirectional flow-based format. 
For each flow, the authors extracted more than 80 attributes and provide additional metadata about IP addresses and attacks. 
Normal user behavior is executed through scripts. 
The data set contains a wide range of attack types like SSH brute force, heartbleed, botnet, DoS, DDoS, web and infiltration attacks. 
CICIDS 2017 is publicly available\footnote{\url{http://www.unb.ca/cic/datasets/ids-2017.html}}.

\textbf{CIDDS-001 \cite{ring2017flow}}.
The CIDDS-001 data set was captured within an emulated small business environment in 2017, contains four weeks of unidirectional flow-based network traffic, and comes along with a detailed technical report with additional information. 
As special feature, the data set encompasses an external server which was attacked in the internet. In contrast to honeypots, this server was also regularly used by the clients from the emulated environment. 
Normal and malicious user behavior was executed through python scripts which are publicly available on GitHub\footnote{\url{https://github.com/markusring/CIDDS}}. 
These scripts allow an ongoing generation of new data sets and can be used by other researches.
The CIDDS-001 data set is publicly available\footnote{\url{http://www.hs-coburg.de/cidds}} and contains SSH brute force, DoS and port scan attacks as well as several attacks captured from the wild.

\textbf{CIDDS-002 \cite{ring2017creation}}. 
CIDDS-002 is a port scan data set which is created based on the scripts of CIDDS-001. 
The data set contains two weeks of unidirectional flow-based network traffic within an emulated small business environment. 
CIDDS-002 contains normal user behavior as well as a wide range of different port scan attacks. 
A technical report provides additional meta information about the data set where external IP addresses are anonymized. 
The data set is publicly available\footnote{\url{http://www.hs-coburg.de/cidds}}.

\textbf{CDX \cite{sangster2009toward}}. 
Sangster el al. \cite{sangster2009toward} propose a concept to create network-based data sets from network warfare competitions and discuss the advantages and disadvantages of such an approach comprehensively. 
The CDX data set contains network traffic of a four day network warfare competition in 2009.
The traffic is recorded in packet-based format and is publicly available\footnote{\url{https://www.usma.edu/crc/sitepages/datasets.aspx}}. 
CDX contains normal user behaviour as well as several types of attacks. 
An additional plan describes metadata about the network structure and IP addresses, but the individual packets are not labeled. 
Further, host-based log files and warnings from an IDS are available.

\textbf{CTU-13 \cite{garcia2014empirical}}. 
The CTU-13 data set was captured in the year 2013 and is available in three formats: packet, unidirectional flow, and bidirectional flow\footnote{\url{https://mcfp.weebly.com/the-ctu-13-dataset-a-labeled-dataset-with-botnet-normal-and-background-traffic.html}}. 
It was captured in a university network and distinguishes 13 scenarios containing different botnet attacks. 
Additional information about infected hosts is provided at the website. 
Traffic was labeled using a three stage approach. 
In the first stage, all traffic to and from infected hosts is labeled as botnet. 
In the second stage, traffic which matches specific filters is labeled as normal. 
Remaining traffic is labeled as background. 
Consequently, background traffic could be normal or malicious. 
The authors recommend a split of their data set into training and test subsets \cite{garcia2014empirical}.

\textbf{DARPA \cite{lippmann2000evaluating},\cite{lippmann20001999},\cite{mitlincolnlaboratory}}. 
The DARPA 1998/99 data sets are the most popular data sets for intrusion detection and were created at the MIT Lincoln Lab within an emulated network environment. 
The DARPA 1998 and DARPA 1999 data sets contain seven and, respectively, five weeks of network traffic in packet-based format, including various kinds of attacks like DoS, buffer overflow, port scans, or rootkits. 
Additional information as well as download links can be found at the website\footnote{\url{https://www.ll.mit.edu/ideval/docs/index.html}}.
In spite (or because) of their wide distribution, the data sets are often criticized for artificial attack injections or the large amount of redundancy \cite{tavallaee2009analysis, mchough2000}.

\textbf{DDoS 2016 \cite{alkasassbeh2016detecting}}. 
Alkasassbeh et al. \cite{alkasassbeh2016detecting} published a packet-based data set which was created using the network simulator NS2 in 2016. 
Detailed information about the simulated network environment is not available. 
The DDoS 2016 data set focuses on different types of DDoS attacks. 
In addition to normal network traffic, the data set contains four different types of DDoS attacks: UDP flood, smurf, HTTP flood, and SIDDOS. 
The data set contains 2.1 million packets and can be downloaded at researchgate\footnote{\url{https://www.researchgate.net/publication/292967044_Dataset-_Detecting_Distributed_Denial_of_Service_Attacks_Using_Data_Mining_Techniques}}.

\textbf{IRSC \cite{zuech2015new}}. 
The IRSC data set was recorded in 2015, using an inventive approach. 
Real network traffic with normal user behavior and attacks from the internet were captured. 
In addition to that, additional attacks were run manually. 
The IDS SNORT\footnote{\url{https://www.snort.org/}} and manual inspection were used for labeling. 
Since the data set is not publicly available due to privacy concerns, we are not able to fill all properties in Table \ref{tbl:datasets}.

\textbf{ISCX 2012 \cite{shiravi2012toward}}. 
The ISCX data set was created in 2012 by capturing traffic in an emulated network environment over one week. 
The authors used a dynamic approach to generate an intrusion detection data set with normal as well as malicious network behavior. 
So-called $\alpha$ profiles define attack scenarios while $\beta$ profiles characterize normal user behavior like writing e-mails or browsing the web. 
These profiles are used to create a new data set in packet-based and bidirectional flow-based format. 
The dynamic approach allows an ongoing generation of new data sets. 
ISCX can be downloaded at the website\footnote{\url{http://www.unb.ca/cic/datasets/ids.html}} and contains various types of attacks like SSH brute force, DoS or DDoS.

\textbf{ISOT \cite{saad2011detecting}}. 
The ISOT data set was created in 2010 by combining normal network traffic from Traffic Lab at Ericsson Research in Hungary \cite{szabo2008validation} and the Lawrence Berkeley National Lab (LBNL) \cite{pang2005} with malicious network traffic from the French chapter of the honeynet project\footnote{\url{http://honeynet.org/chapters/france}}. 
ISOT was used for detecting P2P botnets \cite{saad2011detecting}.
The resulting data set is publicly available\footnote{\url{https://www.uvic.ca/engineering/ece/isot/datasets/}} and contains 11 GB of packet-based data in pcap format.

\textbf{KDD CUP 99 \cite{kddcup}}. 
KDD CUP 99 is based on the DARPA data set and among the most widespread data sets for intrusion detection. 
Since it is neither in standard packet-, nor in flow-based format, it belongs to category \textit{other}. 
The data set contains basic attributes about TCP connections and high-level attributes like number of failed logins, but no IP addresses. 
KDD CUP 99 encompasses more than 20 different types of attacks (e.g. DoS or buffer overflow) and comes along with an explicit test subset. 
The data set includes 5 million data points and can be downloaded freely\footnote{\url{http://kdd.ics.uci.edu/databases/kddcup99/kddcup99.html}}.

\textbf{Kent 2016 \cite{kent-2015-cyberdata1}, \cite{akent-2015-enterprise-data}}.
This data set was captured over 58 days at the Los Alamos National Laboratory network. 
It contains around 130 million flows of unidirectional flow-based network traffic as well as several host-based log files. 
Network traffic is heavily anonymized for privacy reasons.  
The data set is not labeled and can be downloaded at the website\footnote{\url{https://csr.lanl.gov/data/cyber1/}}.

\textbf{Kyoto 2006+ \cite{song2011statistical}}.
Kyoto 2006+ is a publicly available honeypot data set\footnote{\url{http://www.takakura.com/Kyoto\_data/}} which contains real network traffic, but includes only a small amount and a small range of realistic normal user behavior. 
Kyoto 2006+ is categorized as \textit{other} since the IDS Bro\footnote{\url{https://www.bro.org/}} was used to convert packet-based traffic into a new format called sessions. 
Each session comprises 24 attributes, 14 out of which characterize statistical information inspired by the KDD CUP 99 data set. 
The remaining 10 attributes are typical flow-based attributes like IP addresses (in anonymized form), ports, or duration. 
A label attribute indicates the presence of attacks. 
Data were captured over three years. As a consequence of that unusually long recording period, the data set contains about 93 million sessions.

\textbf{LBNL \cite{pang2005}}. 
Research on intrusion detection data sets often refers to the LBNL data set. 
Thus, for the sake of completeness, this data set is also added to the list. 
The creation of the LNBL data set was mainly motivated by analyzing characteristics of network traffic within enterprise networks, rather than publishing intrusion detection data. 
According to its creators, the data set might still be used as background traffic for security researchers as it contains almost exclusively normal user behavior.
The data set is not labeled, but anonymized for privacy reasons, and contains more than 100 hours of network traffic in packet-based format. 
The data set can be downloaded at the website\footnote{\url{http://icir.org/enterprise-tracing/}}.

\textbf{NDSec-1 \cite{beer2017new}}. 
The NDSec-1 data set is remarkable since it is designed as an attack composition for network security. 
According to the authors, this data set can be reused to salt existing network traffic with attacks using overlay methodologies like \cite{aviv2011challenges}.
NDSec-1 is publicly available on request\footnote{\url{http://www2.hs-fulda.de/NDSec/NDSec-1/}} and was captured in packet-based format in 2016. 
It contains additional syslog and windows event log information. 
The attack composition of NDSec-1 encompasses botnet, brute force attacks (against FTP, HTTP and SSH), DoS (HTTP flooding, SYN flooding, UDP flooding), exploits, port scans, spoofing, and XSS/SQL injection.

\textbf{NGIDS-DS \cite{haider2017generating}}.
The NGIDS-DS data set contains network traffic in packet-based format as well as host-based log files. 
It was generated in an emulated environment, using the IXIA Perfect Storm tool to generate normal user behavior as well as attacks from seven different attack families (e.g. DoS or worms).   
Consequently, the quality of the generated data depends primarily on the IXIA Perfect Storm hardware\footnote{\url{https://www.ixiacom.com/products/perfectstorm}}.
The labeled data set contains approximately 1 million packets and is publicly available\footnote{\url{https://research.unsw.edu.au/people/professor-jiankun-hu}}.

\textbf{NSL-KDD \cite{tavallaee2009analysis}}. 
NSL-KDD enhances the KDD CUP 99. 
A major criticism against the KDD CUP 99 data set is the large amount of redundancy\cite{tavallaee2009analysis}. 
Therefore, the authors of NSL-KDD removed duplicates from the KDD CUP 99 data set and created more sophisticated subsets. 
The resulting data set contains about 150,000 data points and is divided into predefined training and test subsets for intrusion detection methods. 
NSL-KDD uses the same attributes as KDD CUP 99 and belongs to the category \textit{other}. 
Yet, it should be noted that the underlying network traffic of NSL-KDD dates back to the year 1998. 
The data set is publicly available\footnote{\url{http://www.unb.ca/cic/datasets/nsl.html}}.

\textbf{PU-IDS \cite{singh2015reference}}.
The PU-IDS data set is a derivative of the NSL-KDD data set.
The authors developed a generator which extracts statistics of an input data set and uses these statistics to generate new synthetic instances. 
As a consequence, the work of Singh et al. \cite{singh2015reference} could be seen as a traffic generator to create PU-IDS which contains about 200,000 data points and has the same attributes and format as the NSL-KDD data set.    
As  NSL-KDD is based on KDD CUP 1999 which in turn is extracted from DARPA 1998, the year of creation is set to 1998 since the input for the traffic generator was captured back then.

\textbf{PUF \cite{sharma2018}}. 
Recently, Sharma et al. \cite{sharma2018} published the flow-based PUF data set which was captured over three days within a campus network and contains exclusively DNS connections. 
38,120 out of a total of 298,463 unidirectional flows are malicious while the remaining ones reflect normal user activity.  
All flows are labeled using logs of an intrusion prevention system. 
For privacy reasons, IP addresses are removed from the data set. 
The authors intend to make PUF publicly available. % (TODO check if its online right now). 

\textbf{SANTA \cite{wheelus2014session}}. 
The SANTA data set was captured within an ISP environment and contains real network traffic. 
The network traffic is labeled through an exhaustive manual procedure and stored in a so-called session-based format. 
This data format is similar to NetFlow but enriched with additional attributes which are calculated by using information from packet-based data. 
The authors spent much effort on the generation of additional attributes which should enhance intrusion detection methods.
SANTA is not publicly available.

\textbf{SSENet-2011 \cite{vasudevan}}.
SSENet-2011 was captured within an emulated environment over four hours. 
It contains several attacks like DoS or port scans.
Browsing activities of participants generated normal user behavior. 
Each data point is characterized by 24 attributes.
The data set belongs to the category \textit{other} since the tool Tstat was used to extract adjusted data points from packet-based traffic. 
We found no information about public availability.

\textbf{SSENet-2014 \cite{bhattacharya2014ssenet}}.
SSENet-2014 is created by extracting attributes from the packet-based files of SSENet-2011 \cite{vasudevan}. 
Thus, like SSENet-2011, the data set is categorized as \textit{other}. 
The authors extracted 28 attributes for each data point which describe host-based and network-based attributes. 
The created attributes are in line with KDD CUP 1999. 
SSENet-2014 contains 200,000 labeled data points and is divided into a training and test subnet.
SSENet-2014 is the only known data set with a balanced training subset. 
Again, no information on public availability could be found.

\textbf{SSHCure \cite{hofstede2014ssh}}. 
Hofstede et al. \cite{hofstede2014ssh} propose SSHCure, a tool for SSH attack detection. 
To evaluate their work, the authors captured two data sets (each with a period of one month) within a university network. 
The resulting data sets are publicly available\footnote{\url{https://www.simpleweb.org/wiki/index.php}} and contain exclusively SSH network traffic. 
The flow-based network traffic is not directly labeled.
Instead, the authors provide additional host-based logs files which may be used to check if SSH login attempts were successful or not. 

\textbf{TRAbID \cite{viegas2017toward}}. 
Viegas et al. proposed the TRAbID database \cite{viegas2017toward} in 2017. 
This database contains 16 different scenarios for evaluating IDS. 
Each scenario was captured within an emulated environment (1 honeypot server and 100 clients). 
In each scenario, the traffic was captured for a period of 30 minutes and some attacks were executed. 
To label the network traffic, the authors used the IP addresses of the clients. 
All clients were Linux machines.
Some clients exclusively performed attacks while most of the clients exclusively handled normal user requests to the honeypot server. 
Normal user behavior includes HTTP, SMTP, SSH and SNMP traffic while malicious network traffic encompasses port scans and DoS attacks. 
TRAbID is publicly available\footnote{\url{https://secplab.ppgia.pucpr.br/trabid}}.

\textbf{TUIDS \cite{gogoi2012packet}, \cite{bhuyan2015towards}}. 
The labeled TUIDS data set can be divided into three parts: TUIDS Intrusion data set, TUIDS coordinated scan data set and TUIDS DDoS data set.
As the names already indicate, the data sets contain normal user behavior and primarily attacks like port scans or DDoS.  
Data were generated within an emulated environment which contains around 250 clients.
Traffic was captured in packet-based and bidirectional flow-based format. 
Each subset spans a period of seven days and all three subsets contain around 250,000 flows. 
Unfortunately, the link\footnote{\url{http://agnigarh.tezu.ernet.in/~dkb/resources.html}} to the data set in the original publication seems to be outdated. 
However, the authors respond to e-mail requests.

\textbf{Twente \cite{sperotto2009labeled}}. 
Sperotto et al. \cite{sperotto2009labeled} published one of the first flow-based intrusion detection data sets in 2008. 
This data set spans six days of traffic involving a honeypot server which offers web, FTP, and SSH services. 
Due to this approach, the data set contains only network traffic from the honeypot and nearly all flows are malicious without normal user behavior. 
The authors analyzed log files and traffic in packet-based format for labeling the flows of this data set. 
The data set is publicly available\footnote{\url{https://www.simpleweb.org/wiki/index.php}} and IP addresses were removed due to privacy concerns.

\textbf{UGR'16 \cite{macia2018ugr}}. 
UGR'16 is a unidirectional flow-based data set. 
Its focus lies on capturing periodic effects in an ISP environment. 
Thus, it spans a period of four months and contains 16,900 million unidirectional flows. 
IP addresses are anonymized and the flows are labeled as normal, background, or attack. 
The authors explicitly executed several attacks (botnet, DoS, and port scans) within that data set. 
The corresponding flows are labeled as attacks and some other attacks were identified and manually labeled as attack. 
Injected normal user behavior and traffic which matches certain patterns are labeled as normal. 
However, most of the traffic is labeled as background which could be normal or an attack. 
The data set is publicly available\footnote{\url{https://nesg.ugr.es/nesg-ugr16/index.php}}.

\textbf{UNIBS 2009 \cite{gringoli2009gt}}. 
Like LBNL \cite{pang2005}, the UNIBS 2009 data set was not created for intrusion detection. 
Since UNIBS 2009 is referenced in other work, it is still added to the list. 
Gringoli et al. \cite{gringoli2009gt} used the data set to identify applications (e.g. web browsers, Skype or mail clients) based on their flow-based network traffic.  
UNIBS 2009 contains around 79,000 flows without malicious behavior.
Since the labels just describe the application protocols of the flows, network traffic is not categorized as normal or attack. 
Consequently, the property label in the categorization scheme is set to \textit{no}. 
The data set is publicly available\footnote{\url{http://netweb.ing.unibs.it/~ntw/tools/traces/)}}.

\textbf{Unified Host and Network Data Set \cite{turcotte2017unified}}. 
This data set contains host and network-based data which were captured within a real environment, the LANL (Los Alamos National Laboratory) enterprise network. 
For privacy reasons, attributes like IP addresses and timestamps were anoynmized in bidirectional flow-based network traffic files.  
The network traffic was collected for a period of 90 days and has no labels. 
The data set is publicly available\footnote{\url{https://csr.lanl.gov/data/2017.html}}.

\textbf{UNSW-NB15 \cite{moustafa2015unsw}}.
The UNSW-NB15 data set encompasses normal and malicious network traffic in packet-based format which was created using the IXIA Perfect Storm tool in a small emulated environment over 31 hours. 
It contains nine different families of attacks like backdoors, DoS, exploits, fuzzers, or worms.  
The data set is also available in flow-based format with additional attributes. 
UNSW-NB15 comes along with predefined splits for training and test. 
The data set includes 45 distinct IP addresses and is publicly available\footnote{\url{https://www.unsw.adfa.edu.au/unsw-canberra-cyber/cybersecurity/ADFA-NB15-Datasets/}}.

\section{Other Data Sources}
\label{sec:further}
Besides network-based data sets, there are some other data sources for packet-based and flow-based network traffic. 
In the following, we shortly discuss data repositories and traffic generators.

\subsection{Data Repositories}
Besides traditional data sets, several data repositories can be found on the internet. 
Since type and structure of those repositories differ greatly, we abstain from a tabular comparison.
Instead, we give a brief textual overview in alphabetical order.
Repositories have been checked on 26 February 2019 with respect to actuality.

\textbf{AZSecure}\footnote{\url{https://www.azsecure-data.org/other-data.html}}.
AZSecure is a repository of network data at the University of Arizona for use by the research community.
It includes various data sets in pcap, arff and other formats some of which are labeled, while other are not.
AZSecure encompasses, among others, the CTU-13 data set \cite{garcia2014empirical} or the Unified Host and Network Data Set \cite{turcotte2017unified}.
The repository is managed and contains some recent data sets.

\textbf{CAIDA}\footnote{\url{http://www.caida.org/data/overview/}}.
CAIDA collects different types of data sets, with varying degree of availability (public access or on request), and provides a search engine.
Generally, a form needs to be filled out to gain access to some of the public data sets. 
Additionally, most network-based data sets can exclusively be requested through an IMPACT (see below) login since CAIDA supports IMPACT as Data Provider.
The repository is managed and updated with new data.

\textbf{Contagiodump}\footnote{\url{http://contagiodump.blogspot.com/}}.
Contagiodump is a blog about malware dumps. 
There are several posts each year and the last post was on 20th March 2018. 
The website contains, among other things, a collection of pcap files from malware analysis.

\textbf{covert.io}\footnote{\url{http://www.covert.io}}. 
Covert.io is a blog about security and machine learning by Jason Trost. 
The blog maintains different lists of tutorials, GitHub repositories, research papers and other blogs concerning security, big data, and machine learning, but also a collection of various security-based data resources\footnote{\url{http://www.covert.io/data-links/}}.
The latest entry was posted on August 14, 2017 by Jason Trost.

\textbf{DEF CON CTF Archive}\footnote{\url{https://www.defcon.org/html/links/dc-ctf.html}}.
DEF CON is a popular annual hacker convention.
The event includes a capture the flag (CTF) competition where every team has to defend their own network against the other teams whilst simultaneously hacking the opponents' networks.
The competition is typically recorded and available in packet-based format on the website.
Given the nature of the competition, the recorded data almost exclusively contain attack traffic and little normal user behavior. 
The website is current and updated annually with new data from the CTF competitions.

\textbf{IMPACT}\footnote{\url{https://www.impactcybertrust.org/}}.
IMPACT Cyber Trust, formerly known as PREDICT, is a community of data providers, cyber security researchers as well as coordinators. 
IMPACT is administrated and up-to-date. 
A data catalog is given on the site to browse the data sets provided by the community.
The data providers are (among others) DARPA, the MIT Lincoln Laboratory, or the UCSD - Center for Applied Internet Data Analysis (CAIDA).
However, the data sets can only be downloaded with an account that may be requested exclusively by researchers from eight selected countries approved by the US Department of Homeland Security.
As Germany is not among the approved locations, no further statements about the data sets can be made.

\textbf{Internet Traffic Archive}\footnote{\url{http://ita.ee.lbl.gov/html/traces.html}}.
The Internet Traffic Archive is a repository of internet traffic traces sponsored by ACM SIGCOMM.
The list includes four extensively anonymized packet-based traces.
In particular, the payload has been removed, all timestamps are relative to the first packet, and IP addresses have been changed to numerical representations.
The packet-based data sets were captured more than 20 years ago and can be downloaded without restriction.

\textbf{Kaggle}\footnote{\url{https://www.kaggle.com/}}. 
Kaggle is an online platform for sharing and publishing data sets. 
The platform contains security-based data sets like KDD CUP 99 and has a search function. 
It allows registered users also to upload and explore data analysis models.

\textbf{Malware Traffic Analysis}\footnote{\url{http://malware-traffic-analysis.net/}}.
Malware Traffic Analysis is a repository which contains blog posts and exercises related to network traffic analysis, e.g. identifying malicious activities. 
Exercises come along with packet-based network traffic which is indirectly labeled through the provided answers to the exercises.
Downloadable files are secured with a password which can be obtained from the website.
The repository is recent and new blog posts are issued almost daily.

\textbf{Mid-Atlantic CCDC}\footnote{\url{http://maccdc.org/}}.
Similar to DEFCON CTF, MACCDC is an annual competition hosted by the US National CyberWatch Center where the captured packet-based traffic of the competitions is made available.
Teams have to assure that services provided by their network are not interrupted in any way.
Similar to the DEFCON CTF archives, MACCDC data contain almost exclusively attack traffic and little normal user behavior. 
The latest competition took place in 2018.

\textbf{MAWILab\footnote{http://www.fukuda-lab.org/mawilab/}}. 
The MAWILab repository contains a huge amount of network traffic over a long time which is captured at a link between USA and Japan. 
For each day since 2007, the repository contains a 15 minute trace in packet-based format. 
For privacy reasons, IP addresses are anonymized  and packet payloads are omitted.  
The captured network traffic is labeled using different anomaly detection methods \cite{mawilab}.

\textbf{MWS\footnote{\url{https://www.iwsec.org/mws/2018/en.html}}.}
The anti malware engineering workshop (MWS) is an annual workshop about malware in Japan. 
The workshop comes along with several MWS data sets which contain packet-based network data as well as host-based log files. 
However, the data sets are only shared within the MWS community which consists of researches in industry and academia in Japan \cite{Hatada2015empowering}. 
The latest workshop took place in 2018.

\textbf{NETRECSEC}\footnote{\url{http://www.netresec.com/?page=PcapFiles}}.
NETRECSEC maintains a comprehensive list of publicly available pcap files on the internet.
Similar to SecRepo, NETRECSEC refers to many repositories mentioned in this work, but also incorporates additional sources like honeypot dumps or CTF events.
Its up-to-dateness can only be judged indirectly as NETRECSEC also refers to data traces from the year 2018.

\textbf{OpenML}\footnote{\url{https://www.openml.org/home}}. 
OpenML is an update-to-date platform for sharing machine learning data sets. 
It contains also security-based data sets like KDD CUP 99.  
The platform has a search function and comes along with other possibilities like creating scientific tasks.

\textbf{RIPE Data Repository}\footnote{\url{https://labs.ripe.net/datarepository}}.
The RIPE data repository hosts a number of data sets. 
Yet, no new data sets have been included for several years. 
To obtain access, users need to create an account and accept the terms and conditions of the data sets.
The repository also mirrors some data available from the Waikato Internet Traffic Storage (see below).

\textbf{SecRepo}\footnote{\url{http://www.secrepo.com/}}.
SecRepo lists different samples of security related data and is maintained by Mike Sconzo.
The list is divided in the following categories: Network, Malware, System, File, Password, Threat Feeds and Other.
The very detailed list contains references to typical data sets like DARPA, but also to many repositories (e.g. NETRECSEC).
The website was last updated on November 20, 2018.

\textbf{Simple Web}\footnote{\url{https://www.simpleweb.org/wiki/index.php/}}.
Simple Web provides a database collection and information on network management tutorials and software.
The repository includes traces in different formats like packet or flow-based network traffic. 
It is hosted by the University of Twente, maintained by members of the DACS (Design and Analysis of Communication Systems) group, and updated with new results from this group.

\textbf{UMassTraceRepository}\footnote{\url{http://traces.cs.umass.edu/}}.
UMassTraceRepository provides the research community with several traces of network traffic.
Some of these traces have been collected by the suppliers of the archive themselves while others have been donated.
The archive includes 19 packet-based data sets from different sources.
The most recent data sets were captured in 2018. 

\textbf{VAST Challenge}\footnote{\url{http://vacommunity.org/tiki-index.php}}.
The IEEE Visual Analytics Science and Technology (VAST) challenge is an annual contest with the goal of advancing the field of visual analytics through competition.
In some challenges, network traffic data were provided for contest tasks.
For instance, the second mini challenge of the VAST 2011 competition involved an IDS log consisting of packet-based network traffic in pcap format.
A similar setup was used in a follow-up VAST challenge in 2012.
Furthermore, a VAST challenge in 2013 deals with flow-based network traffic.

\textbf{WITS: Waikato Internet Traffic Storage}\footnote{\url{https://wand.net.nz/wits/catalogue.php}}.
This website aims to list all internet traces possessed by the WAND research group.
The data sets are typically available in packet-based format and free to download from the Waikato servers. 
However, the repository has not been updated for a long time.

\subsection{Traffic Generators}
Another source of network traffic for intrusion detection research are traffic generators. 
Traffic generators are models which create synthetic network traffic. 
In most cases, traffic generators use user-defined parameters or extract basic properties of real network traffic to create new synthetic network traffic. 
While data sets and data repositories provide fixed data, traffic generators allow the generation of network traffic which can be adapted to certain network structures.

For instance, the traffic generators FLAME \cite{brauckhoff2008flame} and ID2T \cite{vasilomanolakis2016towards} use real network traffic as input. 
This input traffic should serve as a baseline for normal user behavior.  
Then, FLAME and ID2T add malicious network traffic by editing values of input traffic or by injecting synthetic flows under consideration of typical attack patterns.
Siska et al. \cite{siska2010} present a graph-based flow generator which extracts traffic templates from real network traffic. 
Then, their generator uses these traffic templates in order to create new synthetic flow-based network traffic.
Ring et al. \cite{ring2018gan} adapted GANs for generating synthetic network traffic. 
The authors use Improved Wasserstein Generative Adversarial Networks (WGAN-GP) to create flow-based network traffic. 
The WGAN-GP is trained with real network traffic and learns traffic characteristics. 
After training, the WGAN-GP is able to create new synthetic flow-based network traffic with similar characteristics.
Erlacher and Dressler's traffic generator GENESIDS \cite{erlacher2018} generates HTTP attack traffic based on user defined attack descriptions. 
There are many additional traffic generators which are not discussed here for the sake of brevity.
Besides those traffic generators, there are many other traffic generators which are not discussed here.
Instead, we refer to Moln{\'a}r et al. \cite{molnar2013validate} for an overview of traffic generators.

Brogi et al. \cite{brogi2017sharing} come up with another idea that in some sense resembles traffic generators.
Starting out from the problem of sharing data sets due to privacy concerns, they present Moirai, a framework which allows users to share complete scenarios instead of data sets. 
The idea behind Moirai is to replay attack scenarios in virtual machines such that users can generate data on the fly.

A third approach - which is also categorized into the larger context of traffic generators - are frameworks which support users to label real network traffic. 
Rajasinghe et al. present such a framework called INSecS-DCS \cite{rajasingheinsecs} which captures network traffic at network devices or uses already captured network traffic in pcap files as input. 
Then, INSecS-DCS divides the data stream into time windows, extracts data points with appropriate attributes, and labels the network traffic based on a user-defined attacker IP address list.
Consequently, the focus of INSecS-DCS is on labeling  network traffic and on extracting meaningful attributes.
Aparicio-Navarro et al.~\cite{aparicio2014automatic} present an automatic data set labeling approach using an unsupervised anomaly-based IDS. 
Since no IDS is able to classify each data point to the correct class, the authors take some middle ground to reduce the number of false positives and true negatives. 
The IDS assigns belief values to each data point for the classes normal and attack. 
If the difference between the belief values for these two classes is smaller than a predefined threshold, the data point is removed from the data set. 
This approach increases the quality of the labels, but may discard the most interesting data points of the data set.

\section{Observations and Recommendations}
\label{sec:obs}

Labeled data sets are inevitable for training supervised data mining methods like classification algorithms and helpful for the evaluation of supervised as well as unsupervised data mining methods. 
Consequently, labeled network-based data sets can be used to compare the quality of different NIDS with each other. 
In any case, however, the data sets must be representative to be suitable for those tasks. 
The community is aware of the importance of realistic network-based data, and this survey shows that there are many sources for such data (data sets, data repositories, and traffic generators). 
Furthermore, this work establishes a collection of data set properties as a basis for comparing available data sets and for identifying suitable data sets, given specific evaluation scenarios. 

In the following, we discuss some aspects concerning the use of available data sets and the creation of new data sets. 

%While the next paragraph \textit{Perfect data set} provides a recommendation for users, the remaining paragraphs provide recommendations for the creation of new data sets. 

\paragraph*{Perfect data set}
The ever-increasing number of attack scenarios, accompanied by new and more complex software and network structures, leads to the requirement that data sets should contain up-to-date and real network traffic. 
Since there is no perfect IDS, labeling of data points should be checked manually rather than being done exclusively by an IDS. 
Consequently, the perfect network-based data set is up-to-date, correctly labeled, publicly available, contains real network traffic with all kinds of attacks and normal user behavior as well as payload, and spans a long time.
Such a data set, however, does not exist and will (probably) never be created. 
If privacy concerns could be satisfied and real-world network traffic (in packet-based format) with all kind of attacks could be recorded over a sufficiently long time, accurate labeling of such traffic would be very time-consuming. 
As a consequence, the labeling process would take so much time that the data set is slightly outdated since new attack scenarios appear continuously.
However, several available data sets satisfy some properties of a perfect data set.
Besides, most applications do not require a perfect data set - a data set which satisfies certain properties is often sufficient. 
For instance, there is no need that a data set contains all types of attacks when evaluating a new port scan detection algorithm, or there is no need for complete network configuration when evaluating the security of a specific server. 
Therefore, we hope that this work supports researchers to find the appropriate data set for their specific evaluation scenario.

\paragraph*{Use of several data sets}
As mentioned above, no perfect network-based data set exists.
However, this survey shows that there are several data sets (and other data sources) available for packet- and flow-based network traffic. 
Therefore, we recommend users to evaluate their intrusion detection methods with more than one data set in order to avoid over-fitting to a certain data set, reduce the influence of artificial artifacts of a certain data set, and evaluate their methods in a more general context.
In addition to that, Hofstede et al. \cite{Hofstede2018flow} show that flow-based network traffic differs between lab environments and production networks.   
Therefore, another approach could be to use both, emulated respectively synthetic data sets and real world network traffic to emphasize these points. 

In order to ensure reproducibility for third parties, we recommend evaluating intrusion detection methods with at least one publicly available data set.

Further, we would like to give a general recommendation for the use of the CICIDS 2017, CIDDS-001, UGR'16 and UNSW-NB15 data sets. 
These data sets may be suitable for general evaluation settings. 
CICIDS 2017 and UNSW-NB15 contain a wide range of attack scenarios. 
CIDDS-001 contains detailed metadata for deeper investigations.  
UGR'16 stands out by the huge amount of flows. 
However, it should be considered that this recommendation reflects our personal views. 
The recommendation does \textbf{not} imply that other data sets are inappropriate. 
For instance, we only refrain to include the more widespread CTU-13 and ISCX 2012 data sets in our recommendation due to their increasing age.
Further, other data sets like AWID or Botnet are better suited for certain evaluation scenarios.

\paragraph*{Predefined Subsets}
Furthermore, we want to make a note on the evaluation of anomaly-based NIDS. 
Machine learning and data mining methods often use so-called 10-fold cross-validation \cite{han2011data}. 
This method divides the data set into ten equal-sized subsets. 
One subset is used for testing and the other nine subsets are used for training. This procedure is repeated ten times, such that every subset has been used once for testing.
However, this straight-forward splitting of data sets makes only limited sense for intrusion detection. 
For instance, the port scan data set CIDDS-002 \cite{ring2017creation} contains two weeks of network traffic in flow-based format. 
Each port scan within this data set may cause thousands of flows. 
Using 10-fold cross-validation would lead to the situation that probably some flows of each attack appear in the training data set. 
Thus, attack detection in test data is facilitated and generalization is not properly evaluated. 

In that scenario, it would be better to train on week1 and test on week2 (and vice versa) for the CIDDS-002 data set.
Defining subsets on that approach may also consider the impact of concept drift in network traffic over time.  
Another approach for creating suitable subsets might be to split the whole data set based on traffic characteristics like source IP addresses. 
However, such subsets must be well designed to preserve the basic network structures of the data set.
For instance, a training data set with exclusively source IP addresses which represent clients and no severs would be inappropriate. 

Based on these observations, we recommend creating meaningful training and test splits with respect to the application domain IT security. 
Therefore, benchmark data sets should be published with predefined splits for training and test to facilitate  comparisons of different approaches evaluated on the same data.

\paragraph*{Closer collaboration}
This study shows (see Section~\ref{sec:datasets}) that many data sets have been published in the last few years and the community works on creating new intrusion detection data sets continuously.
Further, the community could benefit from closer collaboration and a single generally accepted platform for sharing intrusion detection data sets without any access restrictions. 
For instance, Cermak et al. \cite{cermaktowards} work on establishing such a platform for sharing intrusion detection data sets. 
Likewise, Ring et al. \cite{ring2017flow} published their scripts for emulating normal user behavior and attacks such that they can be used and improved by third parties. 
A short summary of all mentioned data sets and data repositories can be found at our website~\footnote{\url{http://www.dmir.uni-wuerzburg.de/datasets/nids-ds}} and we intend to update this website with upcoming network-based data sets.

\paragraph*{Standard formats} 
Most network-based intrusion detection approaches require standard input data formats and cannot handle preprocessed data.
Further, it is questionable if data sets from category \textit{other} (Section \ref{sec:other}) can be calculated in real time which may affect their usefulness in NIDS. 
Therefore, we suggest providing network-based data sets in standard packet-based or flow-based formats as they are captured in real network environments. 
Simultaneously, many anomaly-based approaches (e.g., \cite{wang2010new} or \cite{zhang2008random}) achieve high detection rates in data sets from the category \textit{other} which is an indicator that the calculated attributes are promising for intrusion detection.
Therefore, we recommend publishing both, the network-based data sets in a standard format and the scripts for transforming the data sets to other formats.  
Such an approach would have two advantages.
First, users may decide if they want to transfer data sets to other formats and a larger number of researchers could use the corresponding data sets. 
Second, the scripts could also be applied to future data sets.

\paragraph*{Anonymization} 
Anonymization is another important issue since this may complicate the analysis of network-based data sets.
Therefore, it should be carefully evaluated which attributes have to be discarded and which attributes may be published in anonymous form.
Various authors demonstrate the effectiveness of using only small parts of payload. 
For example, Mahoney~\cite{mahoney2003network} proposes an intrusion detection method which uses the first 48 bytes of each packet starting with the IP header. 
The flow exporter YAF~\cite{inacio2010yaf} allows the creation of such attributes by extracting the first $n$ bytes of payload or by calculating the entropy of the payload.
Generally, there are several methods for anonymization. 
For example, Xu et al.~\cite{xu2002prefix} propose a prefix-preserving IP address anonymization technique. 
Tcpmkpub~\cite{pang2006devil} is an anonymization tool for packet-based network traffic which allows the anonymization of some attributes like IP addresses and also computes new values for header checksums. 
We refer to Kelly et al.~\cite{kelly2008survey} for a more comprehensive review of anonymization techniques for network-based data.

\paragraph*{Publication}
We recommend the publication of network-based data sets. 
Only publicly available data sets can be used by third parties and thus serve as a basis for evaluating NIDS. 
Likewise, the quality of data sets can only be checked by third parties if they are publicly available.
Last but not least, we recommend the publication of additional metadata such that third parties are able to analyze the data and their results in more detail.

\section{Summary}
Labeled network-based data sets are necessary for training and evaluating NIDS. 
This paper provides a literature survey of available network-based intrusion detection data sets. 
To this end, standard network-based data formats are analyzed in more detail. 
Further, 15 properties are identified that may be used to assess the suitability of data sets. 
These properties are grouped into five categories: General Information, Nature of the Data, Data Volume, Recording Environment and Evaluation. 

The paper's main contribution is a comprehensive overview of 34 data sets which points out the peculiarities of each data set. 
Thereby a particular focus was placed on attack scenarios within the data sets and their interrelationships.  
In addition, each data set assessed with respect to the properties of the categorization scheme developed in the first step.
This detailed investigation aims to support readers to identify data sets for their purposes. 
The review of data sets shows that the research community has noticed a lack of publicly available network-based data sets and tries to overcome this shortage by publishing a considerable number of data sets over the last few years. 
Since several research groups are active in this area, additional intrusion detection data sets and improvements can be expected soon.

As further sources for network traffic, traffic generators and data repositories are discussed in Section \ref{sec:further}.  
Traffic generators create synthetic network traffic and can be used to create adapted network traffic for specific scenarios. 
Data repositories are collections of different network traces on the internet. 
Compared to the data sets in Section \ref{sec:datasets}, data repositories often provide limited documentation, non-labeled data sets or network traffic of specific scenarios (e.g., exclusively FTP connections). 
However, these data sources should be taken into account when searching for suitable data, especially for specialized scenarios. 
Finally, we discussed some observations and recommendations for the use and generation of network-based intrusion detection data sets. 
We encourage users to evaluate their methods on several data sets to avoid over-fitting to a certain data set and to reduce the influence of artificial artifacts of a certain data set. 
Further, we advocate data sets in standard formats including predefined training and test subsets.
Overall, there probably won't be a perfect data set, but there are many very good data sets available and the community could benefit from closer collaboration.

\section*{Acknowledgment}
M.R. and S.W. were supported by the BayWISS Consortium Digitisation.
S.W. is additionally funded by the Bavarian State Ministry of Science and Arts in the framework of the Centre Digitisation.Bavaria (ZD.B).

\ifCLASSOPTIONcaptionsoff
  \newpage
\fi

\end{document}